\documentstyle[epsf,epsfig,wrapfig,here,12pt]{article}    
\begin{document}           
\baselineskip=0.33333in
\begin{quote} \raggedleft TAUP 2833 - 2006
\end{quote}
\vglue 0.5in
\begin{center}{\bf The Significance of Density in\\
the Structure of Quantum Theories}
\end{center}
\begin{center}E. Comay$^*$
\end{center}

\begin{center}
School of Physics and Astronomy \\
Raymond and Beverly Sackler Faculty of Exact Sciences \\
Tel Aviv University \\
Tel Aviv 69978 \\
Israel
\end{center}
\vglue 0.5in
\vglue 0.5in
\noindent
\vglue 0.2in
\noindent
Abstract:

It is proved that density plays a crucial role in the structure of
quantum field theory. The Dirac and the Klein-Gordon equations are
examined. The results prove that the Dirac equation is consistent
with density related requirements whereas the Klein-Gordon equation
fails to do that. Experimental data support these conclusions.

\newpage
The present work reviews very briefly the first steps taken by
the standard method of constructing a quantum field theory. Then,
the need for a self-consistent expression for density is discussed.
Later, this general analysis is examined for the specific cases of a Dirac
field and of Klein-Gordon (KG) fields. The discussion contains
new results proving the significant role of density in the structure
of quantum field theory. Some concluding remarks follow.

Herein, units where $\hbar = c = 1$ are used. The metric is diagonal
and its entries are (1,-1,-1,-1). Greek indices run from 0 to 3.
The subscript symbol $_{,\mu}$ denotes the partial differentiation
with respect to $x^\mu $. An upper dot denotes a differention
with respect to time. Only one kind of dimension is required
for the system of units used here. Thus, dimensions of a variable
are denoted by an expression of the form $[L^n]$, where the
letter $[L]$, enclosed by square brackets, denotes the unit
of length (and should be distinguished from the Lagrangian $L$).

A standard method of constructing a quantum field theory (see e.g.
[1], Section 11.3) begins with the equation of motion of the specific
field discussed
\begin{equation}
\hat {O} \psi = 0,
\label{eq:FEQ}
\end{equation}
where the operator $\hat {O}$ denotes the field's equation. At this
point, a Lagrangian density ${\mathcal L}$ is defined. This
Lagrangian density yields an expression for the action of the system
\begin{equation}
I = \int {\mathcal L} d^4x.
\label{eq:ACTION}
\end{equation}
${\mathcal L}$ is defined so that an application of the variational
principle to its action reproduces $(\!\!~\ref{eq:FEQ})$.

The Hamiltonian density
can be derived from the Lagrangian density ${\mathcal L}$.
Thus,
\begin{equation}
{\mathcal H} = \dot {\psi} \frac {\partial {\mathcal L}}
{\partial \dot {\psi}} - {\mathcal L}.
\label{eq:HAMDEN}
\end{equation}

A spatial integration of  $(\!\!~\ref{eq:HAMDEN})$
\begin{equation}
H = \int {\mathcal H} d^3x
\label{eq:HAM}
\end{equation}
yields the Hamiltonian for the field equation $(\!\!~\ref{eq:FEQ})$.

An alternative and equivalent procedure can be taken.
In this case, the
Lagrangian $L$ is obtained as the spatial integral of the
Lagrangian density ${\mathcal L}$ and the Hamiltonian is derived
from this Lagrangian. These alternatives are equivalent and, as
shown below, both require a self-consistent expression for density.

These steps provide the basis for other steps taken for accomplishing
the structure of the theory. The objective of this work is to analyze
the physical meaning of the operations that begin with
$(\!\!~\ref{eq:FEQ})$ and end with $(\!\!~\ref{eq:HAM})$.
The structure of $(\!\!~\ref{eq:FEQ})$ can be treated in a mathematical
sense as an eigenfunction/eigenvalue problem. The following
analysis aims to show how expressions obtained along the way
from $(\!\!~\ref{eq:FEQ})$ to $(\!\!~\ref{eq:HAM})$ acquire physical
meaning and physical constraints as well.

The left hand side of $(\!\!~\ref{eq:ACTION})$ is dimensionless.
Therefore, since the dimension of $d^4x$ is $[L^4]$, one concludes
that the dimension of the Lagrangian density
${\mathcal L}$ is $[L^{-4}]$. It follows that
the form of the operator
$\hat {O}$ of $(\!\!~\ref{eq:FEQ})$
boils down to the Lagrangian density and {\em affects
the dimension of the wave function.} Thus, one realizes that the
construction of the Lagrangian density changes the meaning of the
wave function: in $(\!\!~\ref{eq:FEQ})$ it is a complex mathematical
function whereas in the Lagrangian density
it acquires dimensions.
This point is used below in an analysis of two specific cases, the
Dirac field and the Klein-Gordon fields.

Now let us turn to the integral $(\!\!~\ref{eq:HAM})$ where the
Hamiltonian $H$ is obtained from the Hamiltonian density
${\mathcal H}$. For this end, the form of the Lagrangian density
${\mathcal L}$ should be examined. Since the operator $\hat {O}$
is independent of the wave function, one finds from the Euler-Lagrange
equation
\begin{equation}
\frac {\partial }{\partial x^\mu} \frac {\partial \mathcal L}
{\partial \psi _{,\mu}} - \frac {\partial \mathcal L}{\partial \psi} = 0
\label{eq:EULAG}
\end{equation}
that the Lagrangian density is a quadratic (or bilinear) function
of the wave function $\psi$. Evidently, the equation of motion
$(\!\!~\ref{eq:FEQ})$ retains its form if one multiplies the
Lagrangian density ${\mathcal L}$ by a numerical factor. On the other
hand, the Hamiltonian $(\!\!~\ref{eq:HAM})$ represents
energy and, for a given system, it should have a specific eigenvalue.

This problem is settled by means of a normalization procedure where
the wave function $\psi $ is multiplied by a normalization factor
which guarantees that the integral $(\!\!~\ref{eq:HAM})$ takes
the correct value. Thus, there is a need for a physically selfconsistent
expression for density.

Now, the integral of density is a Lorentz scalar, because the particle
is found in all Lorentz frames. Hence, one may take the requirements
for particle density from electrodynamics where an expression for
charge density is readily found (see [2], pp. 69-73).
Thus, in a quantum theory, density
must satisfy the following requirements:
\begin{itemize}
\item[{A.}] The dimension of density is $[L^{-3}]$.
\item[{B.}] Density is the 0-component of a 4-vector $j^\mu$.
\item[{C.}] This 4-vector satisfies the continuity equation
\begin{equation}
j^\mu_{,\mu } = 0.
\label{eqCONTINUITY}
\end{equation}

\end{itemize}

These points are known for a very long time. Here they are
used in an analysis of the Dirac and the KG fields. In
particular, a new aspect of these requirements is shown here. Thus,
it is proved that requirements A-C are necessary, but not
sufficient, conditions for constructing a self-consistent 
expression for density of a quantum
field theory.

Let us begin with an analysis of the Dirac field.
Here, the matter part of the Lagrangian density is (see [1], p. 84)
\begin{equation}
{\mathcal L} = \bar \psi[\gamma ^\mu (i\partial _\mu - eA_\mu) - m]\psi ,
\label{eq:DIRACLD}
\end{equation}
As is well known, a 4-current is defined for the Dirac field
\begin{equation}
j^\mu = \bar \psi\gamma ^\mu \psi .
\label{eq:DIRACJMU}
\end{equation}
This 4-current satisfies requirements A-C. The density of
$(\!\!~\ref{eq:DIRACJMU})$
\begin{equation}
\rho =  \bar \psi\gamma ^0 \psi = \psi ^\dagger \psi
\label{eq:DIRACRHO}
\end{equation}
has been used recently (see [3], Section 2)
in an analysis of the Dirac field. The results are:
\begin{itemize}
\item[{1.}] The conserved 4-current depends on $\psi $ and on the
corresponding $\bar \psi $, and is independent of the external
field $A_\mu$. Hence, one can use the positive definite density
$\psi ^\dagger \psi $ and construct an orthonormal basis for the Hilbert
space of solutions. This basis is not affected by changes of external
quantities.
\item[{2.}] The Dirac Hamiltonian operator
is easily extracted from
the Hamiltonian density and is free of
$\psi,\; \bar {\psi}$ and their derivatives. An examination of
the fundamental quantum mechanical equation
\begin{equation}
H\psi = i\frac {\partial \psi}{\partial t}.
\label{eq:HPSI}
\end{equation}
proves that this property is consistent
with the linearity of quantum mechanics and with the superposition
principle as well.
\item[{3.}] Since the Dirac Lagrangian density is
{\em linear} in the time-derivative
$\partial \psi /\partial t$, the corresponding Hamiltonian
density does not contain derivatives of $\psi $
with respect to time. The same is true for the Hamiltonian
differential operator which is extracted from the Hamiltonian
density. Hence, in the case of a Dirac particle,
the fundamental quantum mechanical relation
$(\!\!~\ref{eq:HPSI})$ takes the standard form of an explicit
first-order partial
differential equation. Here a derivative with respect to time
is equated to an expression which is free of time derivatives.
This property does not hold for Hamiltonians that depend on
time derivative operators.
\item[{4.}] If the Dirac Hamiltonian is substituted into
$(\!\!~\ref{eq:HPSI})$ then one finds that it agrees completely with
the Dirac equation obtained as the
Euler-Lagrange equation of the
Lagrangian density of the Dirac field. This property means that
the Dirac's Euler-Lagrange equation does not impose
additional restrictions on the Hamiltonian's eigenfunctions and on their
corresponding eigenvalues.
\item[{5.}] The term $eA^\mu $ of the Dirac Hamiltonian correctly represents
electromagnetic interactions.
\end{itemize}

These results prove that the construction of the Dirac Hamiltonian
proceeds in a straightforward manner and that self-consistent expressions
are obtained. It is shown below that results of the KG field are
inconsistent with points 1-5 above.

Now let us turn to the KG equation. Here one finds two kinds of fields:
one kind of fields uses complex wave functions and the second kind uses
real wave functions. The former is used for describing charged KG
particles and the latter is used in the Yukawa Lagrangian density.
The discussion begins with the complex fields.

The Lagrangian density of the complex KG fields can be found in
Section 3 of [4] (The following expressions are written in
units where $\hbar = c = 1$)
\begin{equation}
{\mathcal L} = (\phi ^*_{,0} -ieV\phi ^*)(\phi _{,0}+ ieV\phi) -
\sum _{k=1}^3 (\phi _{,k}^* +ieA_k \phi ^*)(\phi _{,k} -ieA_k \phi )
- m^2\phi ^* \phi.
\label{eq:PWLD}
\end{equation}
Here, as usual, the symbol $\phi $ denotes the KG wave function.
$V$ and $A_k$ denote the scalar and the vector potentials, respectively.
Using, methods which have become standard,
the Authors of [4] obtain the Hamiltonian density
\begin{equation}
{\mathcal H} = (\phi ^*_{,0} -ieV\phi ^*)(\phi _{,0} + ieV\phi) +
\sum _{k=1}^3 (\phi _{,k}^* +ieA_k \phi ^*)(\phi _{,k} -ieA_k \phi )
+ m^2\phi ^* \phi.
\label{eq:PWHD}
\end{equation}
A 4-current is obtained for this theory and it is shown that it
satisfies requirements A-C obtained earlier in this work. Thus,
the density of this 4-current is (see eq. (42) therein)
\begin{equation}
\rho = i(\phi ^* \phi _{,0} - \phi ^*_{,0} \phi) - 2eV\phi ^* \phi.
\label{eq:PWRHO}
\end{equation}
and the corresponding 3-current is (see eq. (43) therein)
\begin{equation}
\mbox {\boldmath $j$} = i((\nabla \phi ^*) \phi - \phi ^* \nabla \phi) -
2e\mbox {\boldmath $A$}\phi ^* \phi.
\label{eq:PWJ}
\end{equation}

An examination of the Hamiltonian density $(\!\!~\ref{eq:PWHD})$
reveals an alarming aspect. Thus, $(\!\!~\ref{eq:PWHD})$ contains
time derivatives of the wave function. 
It follows that if a Hamiltonian can
be constructed then the Hamiltonian density is expressed in terms
of the Hamiltonian whereas in $(\!\!~\ref{eq:HAM})$ the
Hamiltonian is expressed in terms of the Hamiltonian density.
Certainly, this is an undesirable situation.
However, it is proved below such a Hamiltonian does not exist.

The problem of extracting a covariant
differential operator for the Hamiltonian
of the complex KG
field is discussed in Section 3 of [3], where it is proved that
this task cannot be accomplished. The proof examines the highest
time derivatives of $\phi ^*,\,\phi$ in
the Hamiltonian density $(\!\!~\ref{eq:PWHD})$
and in the density $(\!\!~\ref{eq:PWRHO})$. For $(\!\!~\ref{eq:PWHD})$
one finds a {\em symmetric} expression $\phi ^*_{,0}\phi_{,0}$ whereas
the density $(\!\!~\ref{eq:PWRHO})$ contains the {\em antisymmetric}
term $\phi ^* \phi _{,0} - \phi ^*_{,0} \phi$. Using self-evident
arguments, one infers from these properties
that {\em there is no covariant
differential operator for the complex KG equation.} This
conclusion is consistent with the contents of the
available literature.

Let us turn now to the problem of constructing a Hamiltonian matrix
of the KG equation. Here one should define a self-consistent
inner product $(\phi^*_i,\phi_j)$
for the Hilbert space and construct an appropriate orthonormal basis.
This basis is used in a calculation of the Hamiltonian matrix elements.
Hence, the density expression $(\!\!~\ref{eq:PWRHO})$ must be used.
It is proved below that such an inner product cannot be constructed
for the complex KG field.

Consider 2 states of a positively charged particle
written in spherical polar coordinates
\begin{equation}
\phi_0(t,r,\theta,\varphi) = e^{i\omega_0 t} f_0(r)Y_{00}(\theta,\varphi),
\label{eq:YLM0}
\end{equation}
\begin{equation}
\phi_1(t,r,\theta,\varphi) = e^{i\omega_1 t} f_1(r)Y_{10}(\theta,\varphi).
\label{eq:YLM1}
\end{equation}
where $Y_{lm}$ are the ordinary spherical harmonics (see [5], pp.
510, 511). The radial functions $f_i(r)$ belong to the lowest energy of the
corresponding angular momentum. Hence, they do not change sign and
$f_i(r) \ge 0$.
Using the expression for density $(\!\!~\ref{eq:PWRHO})$, one examines
the inner product of these functions in the case where the external
potential $V$ vanishes. In this case, one finds that the density is
\begin{equation}
\rho = i(\phi _0^* \phi _{1,0} - \phi ^*_{0,0} \phi _1)
\label{eq:DENSITY0}
\end{equation}
Substituting $(\!\!~\ref{eq:YLM0})$ and $(\!\!~\ref{eq:YLM1})$
into the density $(\!\!~\ref{eq:DENSITY0})$
and performing the integration, one
finds
\begin{equation}
\int (\omega _0 + \omega _1)
\phi_0(t,r,\theta,\varphi) \phi_1(t,r,\theta,\varphi) r^2 sin(\theta)
dr\,d\theta \,d\phi = 0,
\label{eq:Y0Y1}
\end{equation}
where the null result is obtained from the orthogonality of
the spherical harmonics $Y_{00}(\theta,\varphi)$ and
$Y_{10}(\theta,\varphi)$.

Now, let us examine these states in the case where an external 
positively charged particle
moves towards the origin along the z-axis and $z>0$.
Hence, in this case,
the external potential $V$ varies in 
space-time and so does the density
$(\!\!~\ref{eq:PWRHO})$. Substituting the new expression for
the density into the integral $(\!\!~\ref{eq:Y0Y1})$, one finds
after a straightforward calculation that the orthogonality of
$\phi_0$ and $\phi_1$ is destroyed. Indeed, the contribution of the
last term of $(\!\!~\ref{eq:PWRHO})$ to the inner product is
\begin{equation}
U = \int -2e
\phi_0(t,r,\theta,\varphi)V\phi_1(t,r,\theta,\varphi) r^2 sin(\theta)
dr\,d\theta \,d\phi
\label{eq:Y0Y1V}
\end{equation}
Let us examine the integrand at two volume elements defined at points
$P_1(r,\theta ,\varphi )$ and $P_2(r,\pi - \theta ,\varphi )$, where
$\theta < \pi/2$. The product $\phi_0 \phi_1$ changes sign at
$P_1,\,P_2$. On the other hand,
$V(r,\theta ,\varphi ) > V(r,\pi - \theta ,\varphi )$.
Hence, $U\ne 0$ and the inner product is destroyed.

This result proves that it is impossible to construct
a self-consistent inner product for the Hilbert space of 
complex KG functions. It follows that a Hamiltonian matrix cannot
be constructed for this field.

This discussion completes the proof that the complex KG field has no
self-consistent expression for density and that its Hamiltonian
cannot be constructed. Another result is
that requirements A-C are only necessary
conditions for a physically self-consistent expression for density
of a quantum field. Indeed,
the 4-vector whose entries are
$(\!\!~\ref{eq:PWRHO})$ and $(\!\!~\ref{eq:PWJ})$
satisfies requirements A-C (see Section 3 of [4])
but it is physically unacceptable.

Let us turn to the case of the real KG field. Using the results of the
complex KG equation, one concludes that, in this case, there is no
expression for density. Indeed, substituting $\phi^* = \phi$ in
$(\!\!~\ref{eq:PWRHO})$, and remembering that a real KG field
cannot carry charge, one finds that the density of a real KG
field vanishes identically [6].

The foregoing discussion can be used for a derivation of another
discrepancy of the KG equation. Here the dimension of the field
function is examined. Thus, in the Lagrangian density of the Dirac
field $(\!\!~\ref{eq:DIRACLD})$, the dimension of the operator
is $[L^{-1}]$. Hence, since the dimension of the Lagrangian
density is $[L^{-4}]$, one finds that
the dimension of the Dirac field
function is $[L^{-3/2}]$.
On the other hand, the dimension of the operator in the KG Lagrangian
density is $[L^{-2}]$. Hence, the dimension of the KG field function
is $[L^{-1}]$. Therefore, it is concluded that
the nonrelativistic limit of the KG equation
disagrees with the Schroedinger equation, because here $\psi^*\psi$
represent density (see [7], p. 54) and $\psi $ has the dimension
$[L^{-3/2}]$.

An examination of contemporary textbooks on quantum field theory
indicates that, at least in the case of the KG equation, the validity of
a density expression is generally
taken for granted when the Hamiltonian is
derived from the Hamiltonian density
(see e.g. [1], p. 26; [8], pp.21, 22; [9], pp. 16-19 etc.).
This work proves that density plays a significant role in the
structure of quantum theories and that it deserves an appropriate
discussion in textbooks.

The issues of the Dirac and the KG equations has a long history of
debates. In particular, Dirac maintained his opinion stating
that the KG equation has no physical merits (see [8],
pp. 7, 8 and [10]). Other
people have adopted a different opinion and most (if not all)
of contemporary textbooks discuss the KG field as a physically
meaningful field. For the
most of the time elapsed, this controversy was based on pure
theoretical arguments. This situation has changed during the last
decades because new experimental data have been accumulated. Thus,
the KG field function $\phi $ depends on
a single set of space-time coordinates.
Hence, like the Dirac field $\psi $, it describes a
structureless pointlike particle.
Now, experimental data tell us that
unlike Dirac particles (electrons, muons, quarks etc.),
the existence of {\em pointlike}
KG particles has not been established. In particular, it is now
recognized that $\pi $ mesons, which are regarded as the primary
example of a KG particle, contain a pair of quark and antiquark
and are not pointlike
particles.

This state of affairs help people take the right course and seek
for theoretical arguments that explain why Nature is so unkind to the KG
theory.

\newpage
References:
\begin{itemize}

\item[{*}] Email: elic@tauphy.tau.ac.il \\
           Internet site: http://www-nuclear.tau.ac.il/$\sim $elic
\item[{[1]}] J. D. Bjorken and S.D. Drell, {\em Relativistic Quantum
Fields} (McGraw-Hill, New York, 1965).
\item[{[2]}] L. D. Landau and E. M. Lifshitz, {\em The Classical
Theory of Fields} (Pergamon, Oxford, 1975).
\item[{[3]}] E. Comay, Apeiron {\bf 12}, no. 1, 26 (2005).
\item[{[4]}] W. Pauli and V. Weisskopf, Helv. Phys. Acta, {\bf 7}, 709 (1934).
English translation: A. I. Miller {\em Early Quantum
Electrodynamics} (University Press, Cambridge, 1994). pp. 188-205.
\item[{[5]}] A. de-Shalit and I. Talmi, {\em Nuclear Shell Theory}
(Academic, New York, 1963).
\item[{[6]}]  V. B. Berestetskii, E. M. Lifshitz and L. P. Pitaevskii
{\em Quantum Electrodynamica} (Pergamon, Oxford, 1982). (See middle
of p. 42.)
\item[{[7]}] L. D. Landau and E. M. Lifshitz, {\em Quantum Mechanics}
(Pergamon, London, 1959).
\item[{[8]}] S. Weinberg, {\em The Quantum Theory of Fields}
(University Press, Cambridge, 1995). Vol. 1.
\item[{[9]}] M. E. Peskin and D. V. Schroeder, {\em An Introduction to 
Quantum Field Theory} (Addison-Wesley, Reading, Mass., 1995).
\item[{[10]}] P. A. M. Dirac {\em Mathematical Foundations of Quantum
Theory}, Editor A. R. Marlow (Academic, New York, 1978). (See pp. 3,4).

\end{itemize}

\end{document}